\documentclass[twocolumn,pra,aps,showpacs]{revtex4}
\usepackage{bm}
\usepackage{mathrsfs}
\usepackage{amsmath}
\usepackage{amssymb}
\usepackage{graphicx}
\usepackage{amsfonts}
\usepackage{amsthm}
\usepackage{color}
\usepackage{dcolumn}
\usepackage{txfonts}
\newtheorem*{thm}{Theorem}

\newtheorem*{lem}{Lemma}

\begin{document}

\title{Extending quantum control of time-independent systems to time-dependent systems}

\author{Zhen-Yu Wang}

\author{Ren-Bao Liu}
\email{rbliu@cuhk.edu.hk}

\affiliation{Department of Physics, The Chinese University of Hong Hong, Shatin,
N. T., Hong Kong, China}
\begin{abstract}
We establish that if a scheme can control a time-independent system
arbitrarily coupled to a generic finite bath over a short period of
time $T$ with control precision $O(T^{N+1})$, it can also realize
the control with the same order of precision on smoothly
time-dependent systems. This result extends the validity of various
universal dynamical control schemes to arbitrary analytically
time-dependent systems.
\end{abstract}


\maketitle
\section{Introduction}

Quantum systems interact with their environments (or baths). This
results in errors in controlling evolution of a quantum system, such
as decoherence and unwanted dynamics. Inspired by phase-refocusing
techniques in magnetic resonance
spectroscopy~\cite{Mehring_NMR,Schweiger}, various schemes of
quantum dynamical
control~\cite{Viola1998PRA,Ban1998DD,Zanardi99PLA,Viola1999PRL,Viola2003EulerianDD,Wocjan2006PRA,Khodjasteh2005PRLcdd,Uhrig2007UDD,Uhrig2009CUDD,Uhrig2010RUDD,West2010QDD,Wang2010NUDD,Yang2010ReviewDD,Viola1999a,Pasini2009Pulse,Khodjasteh09PRL,Khodjasteh2010,Clausen2010PRL,Bensky2010DDControl,West2010DDgate}
have been developed in the context of quantum information processing
to average out undesired coupling through fast open-loop modulation
on the system evolution. These dynamical control schemes have
advantages of correcting errors without measurement, feedback, or
redundant encoding~\cite{Viola1999PRL}. The simplest one is
dynamical decoupling
(DD)~\cite{Viola1998PRA,Ban1998DD,Zanardi99PLA,Viola1999PRL,Viola2003EulerianDD,Wocjan2006PRA,Khodjasteh2005PRLcdd,Uhrig2007UDD,Uhrig2009CUDD,Uhrig2010RUDD,West2010QDD,Wang2010NUDD,Yang2010ReviewDD},
which aims at preservation of arbitrary states (i.e., quantum memory
or NULL quantum control) by achieving a trivial identity evolution.
Recent
experiments~\cite{Biercuk2009QuantumMemory,Du2009Nature,Warren2009UDDchem,Lange2010Science,Ryan2010PRL,AlvarezPRA2010}
have demonstrated the performance of DD. More general aims are to
implement non-trivial quantum
evolutions~\cite{Viola1999a,Pasini2009Pulse,Khodjasteh09PRL,Khodjasteh2010,Clausen2010PRL,Bensky2010DDControl,West2010DDgate}.
Arbitrarily accurate dynamical control can be achieved using a
concatenated design~\cite{Khodjasteh2010}.

A quantum dynamical control is called universal if it has errors up
to an order in short evolution time $T$ for an arbitrary finite
bath. Most universal
schemes~\cite{Viola1999PRL,Viola2003EulerianDD,Wocjan2006PRA,Khodjasteh2005PRLcdd,Uhrig2007UDD,Uhrig2009CUDD,Uhrig2010RUDD,West2010QDD,Wang2010NUDD,Pasini2009Pulse,Khodjasteh09PRL,Khodjasteh2010}
are designed for time-independent systems and their applicability to
time-dependent systems is unclear, except an explicit
extension~\cite{Pasini2010UDDTime} of Uhrig dynamical decoupling
(UDD)~\cite{Uhrig2007UDD,Yang2008PRL} to analytically time-dependent
systems.

In this paper, we prove a theorem that if a dynamical control has
errors to the $N$th order in short evolution time $T$ for a generic
time-independent system, it will automatically achieve the same
order of precision for analytically time-dependent systems. The
theorem establishes the validity of universal DD in
Refs.~\cite{Viola1999PRL,Viola2003EulerianDD,Wocjan2006PRA,Khodjasteh2005PRLcdd,Uhrig2007UDD,Uhrig2009CUDD,Uhrig2010RUDD,West2010QDD,Wang2010NUDD},
optimized pulses in Ref.~\cite{Pasini2009Pulse}, and dynamical
quantum error correction~\cite{Khodjasteh09PRL,Khodjasteh2010} on
non-equilibrium baths. In addition, it greatly simplifies designing
new universal dynamical control schemes since we just need to work
with time-independent models.

We present the proof in Sec.~\ref{sec:TimeControl} and draw the
conclusion in Sec.~\ref{sec:Conclusions}.

\section{Control of time-dependent systems\label{sec:TimeControl}}
\subsection{Universal control of time-independent systems}

Let us first consider a target system coupled to a bath through a
time-independent Hamiltonian\begin{equation}
H_{SB}=\sum_{\alpha=0}^{D-1}S_{\alpha}\otimes
B_{\alpha},\label{eq:Hsb}\end{equation} where $S_{\alpha}$ and
$B_{\alpha}$ are operators of the system and bath, respectively, and
in particular, $S_{0}\equiv I_{S}$ is the identity operator and
$B_{0}$ is the bath internal interaction. We assume that
$S_{\alpha}$ and $B_{\alpha}$ are bounded in spectrum so that a
perturbative expansion of the system-bath propagator driven by
$H_{SB}$ converges for a short evolution time $T$. Otherwise, the
coupling is generic, that is, the details of $B_{\alpha}$ are
unspecified. In Eq.~(\ref{eq:Hsb}),
$\mathcal{S}\equiv\{S_{\alpha}|\alpha=0,\ldots,D-1\}$ does not have
to be the basis of the full Lie algebra. For example, in the pure
dephasing Hamiltonian of a qubit coupled to a bath,
$\mathcal{S}=\{I,\sigma_z\}$, which only generates a sub-algebra of
a qubit.

Control of the system is implemented by applying a Hamiltonian
$V_{c,T}(t)$ on the system. To realize a desired system evolution
$Q$ (e.g., a quantum gate) over a given duration of time $T$,
$V_{c,T}(t)$ scales with $T$ so that ($\hbar=1$)
\begin{align}
U_{c,T}(t) & \equiv\mathcal{T}\exp\left[-i\int_{0}^{t}V_{c,T}(\tau)d\tau\right]\nonumber \\
 & =\mathcal{T}\exp\left[-i\int_{0}^{t/T}V_{c}(\theta)d\theta\right]\equiv U_{c}(t/T),\end{align}
where $\mathcal{T}$ denotes the time-ordering operator, and
$V_c(\theta)=TV_{c,T}(T\theta)$. We consider the case of perfect
control, that is, $U_{c,T}(T)=U_{c}(1)=Q$ is the desired control of
the system. Under influence of the environment, the system-bath
propagator reads
\begin{equation}
U(T)=\mathcal{T}\exp\left(-i\int_{0}^{T}\left[H_{SB}+V_{c,T}(t)\right]dt\right).
\end{equation} The errors induced by $H_{SB}$ can be isolated in the
interaction picture by writing $U(T)=QU_{E}(T)$, where the error
propagator\begin{align}
U_{E}(T)\equiv & \mathcal{T}\exp\left[-i\int_{0}^{T}U_{c,T}^{\dagger}(t)H_{SB}U_{c,T}(t)dt\right]\nonumber \\
= &
\mathcal{T}\exp\left[-iT\int_{0}^{1}U_{c}^{\dagger}(\theta)H_{SB}U_{c}(\theta)d\theta\right].\end{align}
We suppose that the control $V_{c,T}(t)$ has been designed to
suppress the errors due to $H_{SB}$ up to the $N$th order of the
evolution time $T$, which is assumed short, that is,
\begin{equation}
U_{E}(T)=U_{\Omega}\left[1+O\left(T^{N+1}\right)\right],\label{eq:ErrorUeTn}\end{equation}
where $U_{\Omega}$ is an operator commuting with a certain set of
system operators $\Omega$. A control $V_{c,T}(t)$ is universal if
Eq.~(\ref{eq:ErrorUeTn}) holds for arbitrary time-independent
$B_{\alpha}$. Ref.~\cite{Khodjasteh2010} shows that $V_{c,T}(t)$ can
be designed to achieve Eq.~(\ref{eq:ErrorUeTn}) with arbitrary $N$
and $Q\neq I_{S}$ for a general time-independent model
[Eq.~(\ref{eq:Hsb})]. For the special case of dynamical decoupling,
$Q=I_{S}$ and all the operators in $\Omega$ are preserved; if
$\Omega$ spans the full algebra of the system, $U_{\Omega}$ is a
pure bath operator and any system states (hence quantum
correlations) will be protected~\cite{Wang2010NUDD}.

\subsection{Generalization to time-dependent systems\label{subsec:timeH}}
A time-dependent version of Eq.~(\ref{eq:Hsb}) reads
\begin{equation}
H_{SB}^{\prime}(t)=\sum_{\alpha=0}^{D-1}S_{\alpha}\otimes
B_{\alpha}^{\prime}(t),\label{eq:HsbTimeDependent}\end{equation}
where the generic bath operators are assumed analytic in time:
\begin{equation}
B_{\alpha}^{\prime}(t)=\sum_{p=0}^{\infty}B_{p}^{\prime(\alpha)}\frac{1}{p!}t^{p}.\end{equation}
We also assume that the bath operators $B_{p}^{\prime(\alpha)}$ are
bounded in spectrum. We are to prove the following theorem.

\begin{thm}
\label{thm:Ht} If $V_{c,T}(t)$ realizes Eq.~(\ref{eq:ErrorUeTn}) for
an arbitrary time-independent Hamiltonian in Eq.~(\ref{eq:Hsb}), it
will realize the control with the same order of precision for an
arbitrary time-dependent Hamiltonian in
Eq.~(\ref{eq:HsbTimeDependent}), that is, the system-bath propagator
commutes with the system operator set $\Omega$ up to an error
$O(T^{N+1})$,
\begin{align}
U^{\prime}(T) & \equiv\mathcal{T}\exp\left(-i\int_{0}^{T}\left[H_{SB}^{\prime}(t)+V_{c,T}(t)\right]dt\right)\nonumber \\
 & \equiv QU_{E}^{\prime}(T)=QU_{\Omega}^{\prime}\left[1+O\left(T^{N+1}\right)\right],\end{align}
where $U_{\Omega}^{\prime}$ commutes with the operator set $\Omega$.
\end{thm}

\begin{proof}
We write
\begin{equation}U_{E}(T)=e^{-iB_{0}T}\tilde{U}_{E}(T),\label{eq:UeInterPic}\end{equation}
and
\begin{align}
\tilde{U}_{E}(T)&=\mathcal{T}\exp\left[-iT\int_{0}^{1}\sum_{\alpha=1}^{D-1}U_{c}^{\dagger}(\theta)B_{\alpha}(\theta)U_{c}(\theta)d\theta\right],\\
B_{\alpha}(\theta)&=e^{iB_{0}T\theta}B_{\alpha}e^{-iB_{0}T\theta}=\sum_{k=0}^{\infty}[iB_{0},B_{\alpha}]_{k}\frac{\left(T\theta\right)^{k}}{k!},
\end{align}
with
$[iB_{0},B_{\alpha}]_{k+1}\equiv[iB_{0},[iB_{0},B_{\alpha}]_{k}]$
and $[iB_{0},B_{\alpha}]_{0}\equiv B_{\alpha}$. The perturbative
expansion of $\tilde{U}_{E}(T)$ in short time $T$ reads\begin{align}
\tilde{U}_{E}(T)= &
1+\sum_{n=1}^{\infty}\sum_{\vec{\alpha}=1}^{D-1}\sum_{\vec{p}=0}^{\infty}T^{n+|\vec{p}|}S_{n}^{\vec{\alpha},\vec{p}}\otimes\mathcal{B}_{n}^{\vec{\alpha},\vec{p}},\label{eq:UeExpand}\end{align}
with short-hand notations
$\sum_{\vec{\alpha}=1}^{D-1}\equiv\sum_{\alpha_1=1}^{D-1}\cdots\sum_{\alpha_n=1}^{D-1}$,
$\sum_{\vec{p}=0}^{\infty}\equiv\sum_{p_1=0}^{\infty}\cdots\sum_{p_n=0}^{\infty}$,
and $|\vec{p}|\equiv\sum_{j=1}^{n}p_{j}$, where the bath and system
operators are,
\begin{align}
\mathcal{B}_{n}^{\vec{\alpha},\vec{p}}&\equiv\frac{[iB_{0},B_{\alpha_{1}}]_{p_{1}}}{p_{1}!}\cdots\frac{[iB_{0},B_{\alpha_{n}}]_{p_{n}}}{p_{n}!},\\
S_{n}^{\vec{\alpha},\vec{p}} &\equiv\int_{0}^{1}U_{c}^{\dagger}(\theta_{1})S_{\alpha_{1}}U_{c}(\theta_{1})\theta_{1}^{p_{1}}\int_{0}^{\theta_{1}}U_{c}^{\dagger}(\theta_{2})S_{\alpha_{2}}U_{c}(\theta_{2})\theta_{2}^{p_{2}}\nonumber \\
 & \times\cdots\int_{0}^{\theta_{n-1}}U_{c}^{\dagger}(\theta_{n})S_{\alpha_{n}}U_{c}(\theta_{n})\theta_{n}^{p_{n}}d\theta_{1}d\theta_{2}\cdots d\theta_{n},\end{align}
respectively.

For the time-dependent Hamiltonian $H_{SB}^{\prime}(t)$, the
expansion of
$\tilde{U}_{E}^{\prime}(T)\equiv\left(\mathcal{T}e^{-i\int_{0}^{T}B^{\prime}_{0}(t)dt}\right)^{\dagger}U_{E}^{\prime}(T)$
has a similar form
\begin{equation}
\tilde{U}_{E}^{\prime}(T)=1+\sum_{n=1}^{\infty}\sum_{\vec{\alpha}=1}^{D-1}\sum_{\vec{p}=0}^{\infty}T^{n+|\vec{p}|}S_{n}^{\vec{\alpha},\vec{p}}\otimes\mathcal{B}_{n}^{\prime\vec{\alpha},\vec{p}},\label{eq:UeTExpand}\end{equation}
where
$\mathcal{B}_{n}^{\prime\vec{\alpha},\vec{p}}\equiv(B_{p_{1},0}^{\prime(\alpha_{1})}/p_{1}!)\cdots(B_{p_{n},0}^{\prime(\alpha_{n})}/p_{n}!)$
with $B_{p,0}^{\prime(\alpha)}$ defined by the expansion of
$B_{\alpha}^{\prime}(t)$ in the interaction picture,
$\left(\mathcal{T}e^{-i\int_{0}^{t}B^{\prime}_{0}(s)ds}\right)^{\dagger}B_{\alpha}^{\prime}(t)\left(\mathcal{T}e^{-i\int_{0}^{t}B^{\prime}_{0}(s)ds}\right)\equiv\sum_{p=0}^{\infty}B_{p,0}^{\prime(\alpha)}t^{p}/p!$.

By assumption, $V_{c,T}(t)$ realizes Eq.~(\ref{eq:ErrorUeTn}) for a
generic time-independent $H_{SB}$. In Appendix, we give an explicit
construction of $\{B_{\alpha}\}$, for which the set of bath
operators $\{\mathcal{B}_{n}^{\vec{\alpha},\vec{p}}|n+|\vec{p}|\le
N\}$ is linear independent. Eqs.~(\ref{eq:ErrorUeTn}) and
(\ref{eq:UeInterPic}) indicate that $\tilde{U}_{E}(T)$ commutes with
$\Omega$ up to the $N$th order in $T$. If the bath operators
$\{\mathcal{B}_{n}^{\vec{\alpha},\vec{p}}|n+|\vec{p}|\le N\}$ in
Eq.~(\ref{eq:UeExpand}) are linear independent (non-zero, of
course), $S_{n}^{\vec{\alpha},\vec{p}}$ must commute with the system
operator set $\Omega$ for $n+|\vec{p}|\le N$. Since
Eq.~(\ref{eq:UeTExpand}) is also an expansion of
$S_{n}^{\vec{\alpha},\vec{p}}$, $\tilde{U}_{E}^{\prime}(T)$ and
hence $U_{E}^{\prime}(T)$ commute with the system operator set
$\Omega$ up to an error $O(T^{N+1})$.
\end{proof}

The theorem extends the validity of universal dynamical control to
analytically time-dependent systems. Note that in
Eq.~(\ref{eq:Hsb}), $\mathcal{S}=\{S_{\alpha}|\alpha=0,\ldots,D-1\}$
does not have to span the full algebra of the system but should
contain the identity operator $I_{S}$. Therefore the theorem does
not rely on the specific algebra generated by $\mathcal{S}$ and is
general for any system-bath interactions provided that the total
generic Hamiltonian includes a free bath term $I_{S}\otimes B_{0}$.
It should be stressed that because the details of the bath operators
$B^{\prime}_{\alpha}(t)$ are unspecified, the dynamical control is
still valid if we do the following variation:
\begin{equation} H_{SB}^{\prime}(t)\rightarrow H_{S}(t)\otimes
I_{B}+\sum_{\alpha=0}^{D-1}S_{\alpha}\otimes
B_{\alpha}^{\prime}(t),\label{eq:HsbExtend}\end{equation} where
$H_{S}(t)$ is a bounded system operator in the space spanned by
${\mathcal S}$ and analytic in time and $I_{B}$ is the identity
operator of the bath. Actually, the first term in the right-hand
side of Eq.~(\ref{eq:HsbExtend}) can be readily absorbed into the
system-bath coupling. The drift errors introduced by the system's
internal Hamiltonian $H_{S}(t)$ are also eliminated.

The operator $H_{SB}^{\prime}(t)$ is required to be bounded for any
$t\in[0,T]$. However, to implement a system evolution
$U_{c,T}(T)=U_{c}(1)=Q\neq I_{S}$, the control $V_{c,T}(t)$ must
scale as $\sim1/T$; this scaling induces a faster evolution on the
system as $T\rightarrow0$ in the limit of instantaneous pulses and
this is the requirement of any dynamical control schemes to suppress
errors: the evolution of the system driven by the control field
$V_{c,T}(t)$ needs to be faster than the bath evolution induced by
$H_{SB}$ or $H_{SB}^{\prime}(t)$.

\section{Conclusions\label{sec:Conclusions}}

We have proved that a universal dynamical control which implements a
quantum evolution of a system up to an error $O(T^{N+1})$ in total
evolution time $T$ for a generic time-independent system-bath
Hamiltonian automatically suppresses errors to $O(T^{N+1})$ for any
analytically time-dependent Hamiltonians. The extension of various
universal dynamical control
schemes~\cite{Viola1999PRL,Viola2003EulerianDD,Wocjan2006PRA,Khodjasteh2005PRLcdd,Uhrig2007UDD,Uhrig2009CUDD,Uhrig2010RUDD,West2010QDD,Wang2010NUDD,Pasini2009Pulse,Khodjasteh09PRL,Khodjasteh2010}
to arbitrary analytically time-dependent systems is therefore
established. This result also simplifies the design of other
universal control schemes.  The current research raises an
interesting question for future study: Are there minimal models for
which a control scheme works with a certain degree of precision will
work with the same order of precision for arbitrary systems,
time-independent or not?

\begin{acknowledgments}
This work was supported by Hong Kong GRF CUHK402209, the CUHK
Focused Investments Scheme, and National Natural Science Foundation
of China Project 11028510.
\end{acknowledgments}

\appendix*
\section{}
Here we give an explicit construction of $\{B_{\alpha}\}$, so that
the set of bath operators
$\{\mathcal{B}_{n}^{\vec{\alpha},\vec{p}}|n+|\vec{p}|\le N\}$ in
Eq.~(\ref{eq:UeExpand}) are linear independent. For that purpose we
prove the following lemma.
\begin{lem}
For a finite number $R$, there exists a construction of $K^{\prime}$
Hermitian operators $\{O_{k}\}_{k=1}^{K^{\prime}}$, such that all
the operator products $O_{k_{1}}O_{k_{2}}\cdots O_{k_{r_{j}}}$ for
different sequences $(k_{1}k_{2}\cdots k_{r_{j}})$ with $1\leq
k_{l}\leq K^{\prime}$ and $1\leq r_{j}\leq R$ are linear
independent.
\end{lem}
\begin{proof}Let $K=K^{\prime}+1$ and $\{|l\rangle|l=0,\ldots,(K^{R}+1)K\}$
be an orthonormal basis. We construct the following Hermitian
operators in the Hilbert space expanded by this
basis
\begin{equation} O_{k}=\sum_{l=0}^{K^{R}}|l\rangle\langle
Kl+k|+\text{h.c.}.\end{equation} Examination of the projection
$|0\rangle\langle0|O_{k_{1}}O_{k_{2}}\cdots O_{k_{r_{j}}}$ shows
that the operator $O_{k_{1}}O_{k_{2}}\cdots O_{k_{r_{j}}}$ contains
one and only one component of the form $|0\rangle\langle l|$,
explicitly, $|0\rangle\langle(k_{1}k_{2}\cdots k_{r_{j}})_{K}|$,
where
\begin{equation} (k_{1}k_{2}\cdots k_{r_{j}})_{K}\equiv
K^{r_{j}-1}k_{1}+\cdots+Kk_{r_{j}-1}+k_{r_{j}},\end{equation} is a
number of base $K$ with $1\leq k_{l}\leq K-1$. Therefore all the
operator products $O_{k_{1}}O_{k_{2}}\cdots O_{k_{r_{j}}}$ for
different sequences $(k_{1}k_{2}\cdots k_{r_{j}})$ with $1\leq
k_{l}\leq K^{\prime}$ and $1\leq r_{j}\leq R$ are linear
independent.
\end{proof}
An explicit construction of $\{B_{\alpha}\}$ reads
\begin{subequations}
\begin{align}
B_{\alpha} &=\sum_{r=0}^{N-1}|r\rangle\langle r|\otimes
h_{r}^{(\alpha)},\text{ for }\alpha\geq1,\\
B_{0} &=\sum_{r=0}^{N-1}|r\rangle\langle r+1|\otimes
I_{h}+\text{h.c.},
\end{align}
\end{subequations}
where $\{|r\rangle|r=0,\ldots,N-1\}$ is an $N$-dimensional
orthonormal basis with the periodic condition
$|r+N\rangle=|r\rangle$, $h_{r}^{(\alpha)}$ is an Hermitian
operator, and $I_{h}$ is the identity operator. Using the Lemma, we
have a construction of the operators $\{h_{r}^{(\alpha)}|0\leq r\leq
N-1,1\leq\alpha\le D-1\}$ such that all operator products
$h_{r_{1}}^{(\alpha_{1})}\cdots h_{r_{n}}^{(\alpha_{n})}$ are linear
independent for different sequences of
$(\alpha_{1},\ldots,\alpha_{n})$ and $(r_{1},\ldots,r_{n})$. We
decompose $B_{0}=B_{+}+B_{-}$ with
$B_{+}\equiv\sum_{r=0}^{N-1}|r\rangle\langle r+1|\otimes I_{h}$ and
$B_{-}=(B_{+})^{\dag}$. Some calculation gives
\begin{equation}
\frac{[iB_{+},B_{\alpha}]_{p}}{p!}=\sum_{r=0}^{N-1}|r\rangle\langle
r+p|\otimes\left(\sum_{k=0}^{p}c^{(k)}_{p}h_{r+p-k}^{(\alpha)}\right),
\end{equation}
where $p\geq0$ and $c^{(k)}_{p}=(-1)^{k}i^{p}/[k!(p-k)!]$ is a
non-zero coefficient. Therefore
\begin{align}
&\left\langle0\left|\mathcal{B}_{n}^{\vec{\alpha},\vec{p}}\right||\vec{p}|\right\rangle= \left\langle0\left|\frac{[iB_{+},B_{\alpha_{1}}]_{p_{1}}}{p_{1}!}\cdots\frac{[iB_{+},B_{\alpha_{n}}]_{p_{n}}}{p_{n}!}\right||\vec{p}|\right\rangle\nonumber \\
&=
\left(\sum_{k=0}^{p_{1}}c^{(k)}_{p_{1}}h_{p_{1}-k}^{(\alpha_{1})}\right)\left(\sum_{k=0}^{p_{2}}c^{(k)}_{p_{2}}h_{p_{1}+p_{2}-k}^{(\alpha_{2})}\right)
 \cdots\left(\sum_{k=0}^{p_{n}}c^{(k)}_{p_{n}}h_{p_{1}+\cdots+p_{n}-k}^{(\alpha_{n})}\right).\label{eq:B0pExpand}
 \end{align}
There is one and only one operator product
$h_{p_{1}}^{(\alpha_{1})}h_{p_{1}+p_{2}}^{(\alpha_{2})}\cdots
h_{p_{1}+\cdots+p_{n}}^{(\alpha_{n})}$ in
$\left\langle0\left|\mathcal{B}_{n}^{\vec{\alpha},\vec{p}}\right||\vec{p}|\right\rangle$.
If $n+|\vec{p}|\le N$, the operators
$h_{p_{1}}^{(\alpha_{1})}h_{p_{1}+p_{2}}^{(\alpha_{2})}\cdots
h_{p_{1}+\cdots+p_{n}}^{(\alpha_{n})}$ are linear independent for
different $\{n,\vec{\alpha},\vec{p}\}$ according to the Lemma. Thus
for $n+|\vec{p}|\le N$, $n\geq1$, and $|\vec{p}|\geq0$,
$\left\langle0\left|\mathcal{B}_{n}^{\vec{\alpha},\vec{p}}\right||\vec{p}|\right\rangle$
and hence $\mathcal{B}_{n}^{\vec{\alpha},\vec{p}}$ are linear
independent for different $\{n,\vec{\alpha},\vec{p}\}$.


\end{document}